\documentclass[12pt,reprint,twocolumn,nofootinbib]{revtex4-1}
\pdfoutput=1
\usepackage[utf8]{inputenc}
\usepackage[english]{babel}
\usepackage{physics}
\usepackage{graphicx}
\usepackage{amsmath,amssymb,amsfonts}
\usepackage{float}
\usepackage[caption=false]{subfig}
\usepackage{xcolor}
\newcommand{\revisionA}[1]{\textcolor{black}{\textrm{#1}}} 
\newcommand{\revisionB}[1]{\textcolor{black}{\textrm{#1}}} 
\newcommand{\revisionC}[1]{\textcolor{black}{\textrm{#1}}} 
\newcommand{\textswap}{\textsc{swap}} 

\begin{document}
\title{State-dependent Routing Dynamics in Noisy Quantum Computing Devices}
\author{Ronald J. Sadlier}
\affiliation{Bredesen Center for Interdisciplinary Research and Graduate Education, University of Tennessee, Knoxville, USA}
\affiliation{Quantum Computing Institute, Oak Ridge National Laboratory, Oak Ridge, Tennessee USA}
\author{Travis S. Humble}
\email{humblets@ornl.gov}
\thanks{This manuscript has been authored by UT-Battelle, LLC under Contract No.~DE-AC05-00OR22725 with the U.S. Department of Energy. The United States Government retains and the publisher, by accepting the article for publication, acknowledges that the United States Government retains a non-exclusive, paid-up, irrevocable, world-wide license to publish or reproduce the published form of this manuscript, or allow others to do so, for United States Government purposes. The Department of Energy will provide public access to these results of federally sponsored research in accordance with the DOE Public Access Plan. (http://energy.gov/downloads/doe-public-access-plan).}
\affiliation{Bredesen Center for Interdisciplinary Research and Graduate Education, University of Tennessee, Knoxville, USA}
\affiliation{Quantum Computing Institute, Oak Ridge National Laboratory, Oak Ridge, Tennessee USA}
\begin{abstract}
Routing plays an important role in programming noisy, intermediate-scale quantum (NISQ) devices, where limited connectivity in the register is overcome by swapping quantum information between locations. However, routing a quantum state using noisy gates introduces non-trivial noise dynamics, and deciding on an optimal route to minimize accumulated error  requires estimates of the expected state fidelity. Here we validate a model for state-dependent routing dynamics in a NISQ processor based on correlated binary noise. We develop a composable, state-dependent noise model for \textsc{cnot} and \textsc{swap} operations that can be characterized efficiently using pair-wise experimental measurements, and we compare model predictions with tomographic state reconstructions recovered from a quantum device. These results capture the state-dependent routing dynamics that are needed to guide routing decisions for near-real time operation of NISQ devices.
\end{abstract}
\maketitle
\section{Introduction}
\begin{figure}[tbh]
\centering
\includegraphics[width=0.9\linewidth]{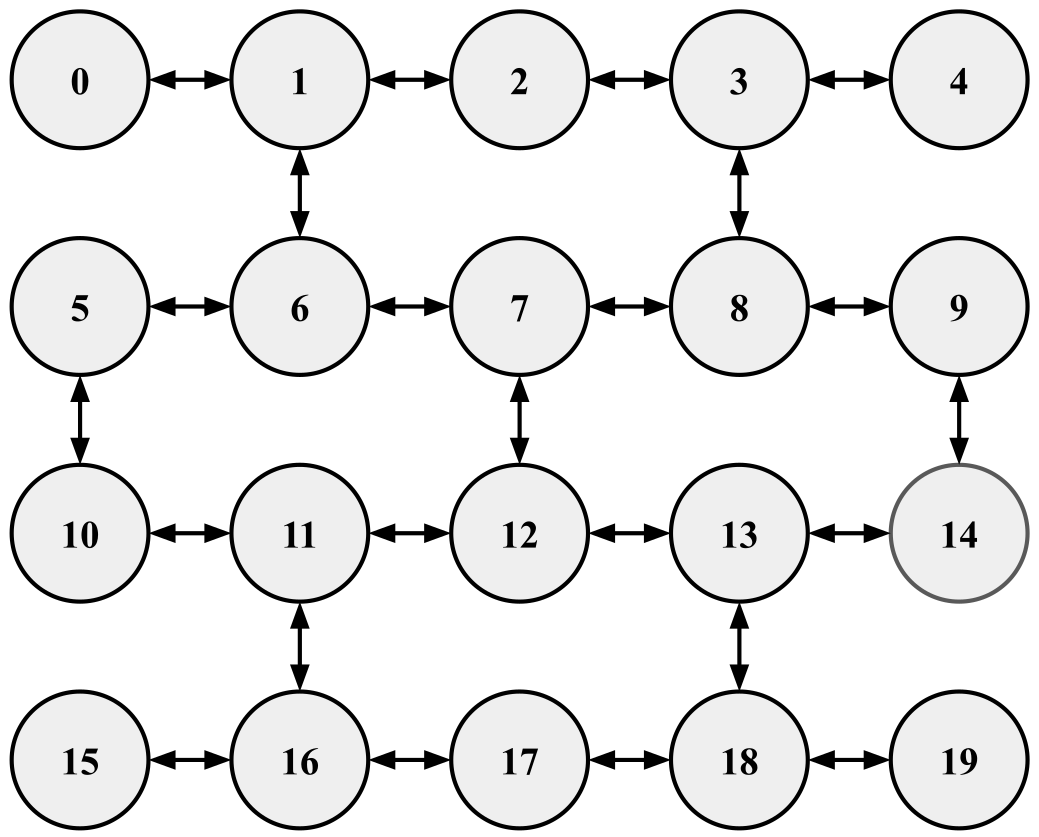}
\caption{Routing a quantum state through the hexagonal topology of the 20-qubit \textit{boeblingen} device from IBM swaps information stored in the register elements (circles) using the available two-qubit gates (edges). Note that edges are directional due to asymmetries in the gate errors.}
\label{fig:boeblingen_graph}
\end{figure}

Programming a quantum computer produces a sequence of executable instructions that respect the logical control of the algorithm as well as the physical constraints of the device \cite{haner2018software,venturelli2018compiling,mccaskey2020xacc,smith2020open}. In current noisy intermediate-scale quantum (NISQ) processors, leading constraints include the modest size and limited connectivity of the quantum register and the limitations on the available instruction set architecture  \cite{britt2017instruction,butko2019understanding}. For example, as shown in Fig.~\ref{fig:boeblingen_graph}, a schematic of the recent IBM \textit{boeblingen} processor indicates 20 available register elements constrained to a hexagonal lattice connected by noisy, directional gate operations. 
\par 
Routing is a method to move quantum information from a source register element to a destination register elements. Routing decisions account for the register connectivity to satisfy simultaneously the logical requirements of a program and the physical constraints of a device \cite{childs_et_al:LIPIcs:2019:10395,9256696}. By exchanging the quantum state stored in connected register elements using a series of logical \textsc{swap} operations, routing  manipulates storage of the quantum state. 
The simplest route uses the shortest sequence of noiseless \textsc{swap} operations. However, the noisy registers and erroneous gate operations in NISQ devices introduce non-trivial changes to the quantum state and optimal routing decisions may therefore be designed to maximize the fidelity
\begin{equation}
\label{eq:fidelity}
F(\ket{\Psi},\rho) = \sqrt{\bra{\Psi}\rho\ket{\Psi}}
\end{equation}
between an ideal routed state $\Psi$ and the potentially noisy, mixed state $\rho$ that represents the actual state of the register. In general, selection of the optimal route requires characterization of the device noise, which in turn depends on a model for the register noise and gate errors. 
\par 
Recently, a variety of heuristics methods for routing have been tailored to NISQ devices, whereby selecting optimal routes relies on modeling the effects of gate noise on the quantum state \cite{Tannu:2019:QCE:3297858.3304007, Murali:2019:NCM:3297858.3304075,vanmeter2019,10.1117/12.2526670,o2019generalized,lao2019mapping}. These heuristic methods estimate the influence of gate noise on either routed state fidelity or expected probability of error. For example, Tannu and Querrishi examined noise-aware routing methods that account for  variations in gate errors while minimizing the total probability of failure \cite{Tannu:2019:QCE:3297858.3304007}.  Similarly, Murali et al.~developed methods for compiling noisy gate sequences a series of scheduling and routing constraints cast as a multi-objective optimization problem \cite{Murali:2019:NCM:3297858.3304075}. Nishio et al.~developed estimated success probability using error rates derived randomized benchmarking to establish a composable model for routing fidelity \cite{vanmeter2019}. Sadlier and Humble presented routing selection as minimization of the shortest weighted path using characterization of the underlying \textsc{swap} and \textsc{cnot} operations \cite{10.1117/12.2526670}. O'Gorman et al.~addressed multi-qubit routing using general swap networks based on routing via matchings \cite{o2019generalized}, while Lao et al.~accounted for a broad set of constraints in the mapping and subsequently, routing, problem for superconducting technologies \cite{lao2019mapping}. 
\par 
A critical consideration for constructing such routing heuristics is the nature of the underlying noise. While current models account for the observed spatial variability in gate noise, it is also known that the errors induced by a gate may depend on the quantum state itself \cite{rudinger2019probing,tannu2019mitigating}. State-dependent noise arises from asymmetry in the underlying noise source \cite{niu2019learning} and represent a refinement of conventional, state-independent models based on randomized benchmarking estimates of error rate \cite{Tannu:2019:QCE:3297858.3304007, Murali:2019:NCM:3297858.3304075,vanmeter2019,10.1117/12.2526670}. 
\par 
Here we develop a state-dependent noise model to describe the observed dynamics that arise during routing within a current NISQ device. Using tomographic reconstructions of experimentally routed quantum states, we quantify differences in the observed and expected fidelities using various starting states. We develop a state-dependent noise model to account for this behavior, and we subsequently estimate parameters for this model using experimental characterizations of a NISQ device. The resulting composable, state-dependent error model requires a relatively small number of characterization circuits and is easily incorporated into a heuristic for estimating routed-state fidelities. We compare the estimated fidelities obtained using this model with the tomographic results that probe the actual routed state. We confirm that the heuristic noise  captures the relative difference in behavior for routes of varying lengths across the register.
\par 
The remainder of the presentation is organized as follows: in Sec.~\ref{sec:route}, we present an overview of quantum state routing, which includes the metrics used for characterizing routing performance. In Sec.~\ref{sec:model}, we present a model of how state-dependent noisy gate operations influence routing performance. We discuss how to estimate parameters for this model using characterization of the \textsc{swap} and \textsc{cnot} gate operations as well as how to compose these noisy gate models to describe larger \textsc{swap} programs. In Sec.~\ref{sec:test}, we present results from testing and evaluation of the state-dependent routing dynamics on a superconducting transmon quantum processor provided by IBM, and we conclude with a discussion of these results in Sec.~\ref{sec:disc}.
\section{Routing on Quantum Processors}
\label{sec:route}
The task of finding the optimal route of a state over a quantum register is constrained by the limited connectivity between register elements and the noise present in the gate operations need to move quantum information. In general, the  optimal path cannot be efficiently estimated as this requires a priori information about the expected register state prepared by a quantum program. Instead, heuristics are necessary to predict the effects of gate noise on an arbitrary quantum state. Sadlier and Humble have shown previously that the transition probabilities characterizing individual \textsc{cnot} and \textsc{swap} gates can be used to estimate the optimal path as a shortest weighted path problem \cite{10.1117/12.2526670}. However, that approach depends on the underlying noise model and raises questions about the most accurate model for current NISQ hardware. Experimental characterizations reported previously indicate that a noise model which accounts for the state of the routed state is need to improve routing decisions \cite{10.1117/12.2526670}. When experimental transition probabilities demonstrate clear state dependence, a state-dependent noise model is needed to account for these routing dynamics.
\subsection{\textsc{SWAP} Program}
A \textsc{swap} program moves the quantum state stored in a source register $q_0$ to a destination register $q_{n-1}$ while satisfying the physical constraints imposed by the addressability of \textsc{swap} gates between $n$ register elements, i.e, the connectivity. 
Denoting an idealized \textsc{swap} instruction acting on register elements $q_i$ and $q_j$ at time $t_{\ell}$ as $\textsc{swap}^{(\ell)}_{(i,j)}$, the gate logic is defined by the unitary transformation
\begin{equation}
    U^{(s)}_{(i,j)}\ket{\phi_{i},\psi_{j}} =\ket{\psi_{i},\phi_{j}}
\end{equation}
for arbitrary states $\phi$ and $\psi$. The gate is linear and applies to superpositions of such states as well. In modeling a \textswap~program under ideal conditions, we assume the complete register is initialized in the separable pure state
\begin{equation}
    \ket{\Psi(t_0)} = \ket{\psi_0}\otimes\ket{\varphi_1}\ldots\ket{0_{n-2}}\otimes\ket{0_{n-1}}
\end{equation}
where $\psi$ and $\varphi$ label the initial quantum states for the elements $q_0$ and $q_1$ in a register of size $n$ at the initial time $t_0$. A \textsc{swap} program consisting of $N$ operations moves the information from element $q_0$ to element $q_{n-1}$ in a time $t_N$. This program produces the ideal register state
\begin{align}
    \ket{\Psi(t_N)} & = \prod_{\ell=1}^{N} { U^{(s)}_{p_\ell} \ket{\Psi(t_{0})} } \\ 
    & =  \ket{\varphi_0}\otimes\ket{0_1}\ldots\ket{0_{n-2}}\otimes\ket{\psi_{n-1}}
\end{align}
where $p_{\ell}$ is the $\ell$-th pair of qubits addressed by a \textsc{swap} gate. A sequence $(p_{1}, p_{2}, \ldots, p_{N})$ that defines the above program must satisfy the connectivity constraints of the underlying register.
\par
As NISQ devices fail to realize the above idealized logical states, we model the quantum state prepared by a \textsc{swap} program in the terms of a density matrix with the initial value given as
\begin{equation}
    \rho_0 = \ket{\Psi(t_0)}\bra{\Psi(t_0)}
\end{equation}
and intermediate values defined by
\begin{equation}
\label{eq:karl}
    \rho_{\ell} = \xi^{(\ell)}(\rho_{\ell-1}) 
\end{equation}
where we use the channel operator $\xi^{(\ell)}$ acting on a density matrix $\rho_{\ell-1}$ to describe noisy transformations for $\ell = 1$ to $N$. In general, the channel operator may model interactions between all registers as well as the environment. However, we will only use $\xi^{(\ell)}$ to model the noisy interactions between the pair of registers $p_{\ell}$ that corresponds with the $\ell$-th operation. Further details on the channel model are presented below. 
\par
We use the fidelity defined in Eq.~(\ref{eq:fidelity}) as a measure of the distance between two quantum states to evaluate the probability of success of a \textsc{swap} program. 
For a given target state $\ket{\Psi(t_N)}$, the fidelity will characterize the accuracy by which the state $\rho_{N}$ is prepared by the \textsc{swap} program. In our analysis, we investigate the fidelity of the prepared final state $\rho_{N}$ relative to the ideal final state $\Psi(t_N)$.
\section{Modeling Noisy Quantum Gates}
\label{sec:model}
In this section, we describe a correlated binary model for noisy \textsc{cnot} and textsc{swap} gates based on the channel operator expression given by Eq.~(\ref{eq:karl}). This noise model accounts for state-dependent error while being easily composable under sequential applications. The binary limitation of the model also make it easily characterized experimentally. We first present the noise model and composition rule before describing its application to noisy \textsc{cnot} and \textsc{swap} gates. 
\subsection{State-dependent Binary Noise}
Consider an initial density matrix $\rho_{0}$ acted upon by a channel operator $\xi_{G}$, which describes the mixed state prepared by a noisy realization of the unitary gate $G$ as
\begin{equation}
    \rho_{1} = \xi_{G}(\rho_0)
\end{equation}
where we expand the channel in an operator-sum representation 
\begin{equation}
    \xi_{G}(\rho) = \sum_{k}{E_{k}^{(G)} U_{G} \rho U_{G}^{\dagger} E_{k}^{(G)\dagger}}
\end{equation}
with $U_{G}$ defining the noiseless gate operation. This expression models gate noise using the operators $E_{k}^{(G)}$ as acting after the ideal gate has been applied. The noise operator decomposition must satisfy the completeness relation
\begin{equation}
    \sum_{k} {E_{k}^{(G)\dagger} E_{k}^{(G)}} = 1
\end{equation}
where selection of the operators are tuned to characterize the noise of interest. 
\par
\begin{figure}[t]
\centering
\includegraphics[width=\linewidth]{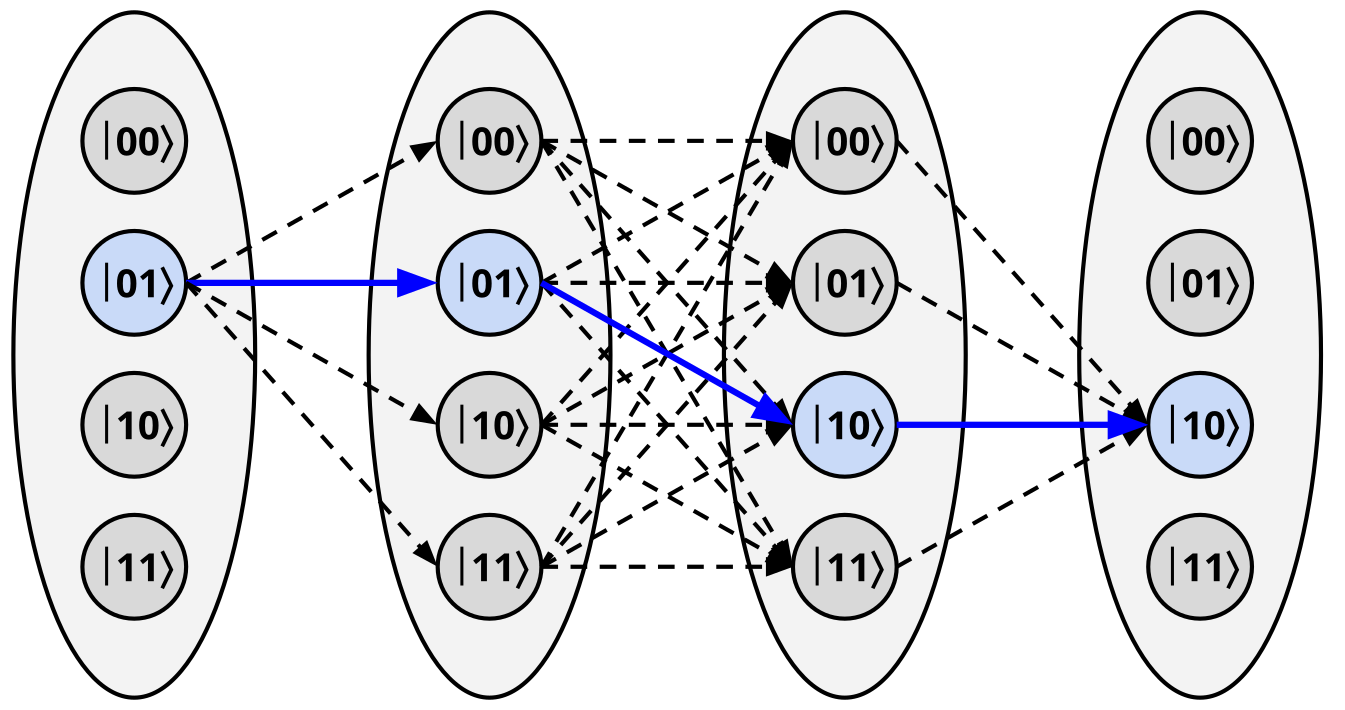}
\caption{
The binary noise channel for a two-qubit gate may be interpreted as a four-state Markov chain. In this example of three successive gates, the blue lines indicate the expected transitions between states, while dashed lines indicate all of the error pathways leading to the expected output state.
}
\label{fig:cnot_state_diagram}
\end{figure}

We consider the case of binary noise to characterize two-qubit gate operations. Noise operators are modeled as products of the identity and $X$ operators. This noise process corresponds to the classical four-state Markov chain shown in Fig.~\ref{fig:cnot_state_diagram}, which highlights transitions between computational basis states during successive noisy operations. We use the binary expansions $j = j_1 + 2j_2$ and $k = k_1 + 2k_2$ with $j_i, k_i \in \{0,1\}$ to define each noise operator, e.g., as
\begin{equation}
\label{eq:binmod}
    E_{k}^{(G)} =  X_{1}^{k_1} X_2^{k_2} \hat{q}_{k}^{(G)}
\end{equation}
where the amplitude operator $\hat{q}_{k}^{(G)}$ determines a state-dependent transition probability. We restrict the state-dependent amplitude operator to be diagonal in the computational basis, defined as 
\begin{equation}
\hat{q}_{k}^{(G)}\ket{j_1, j_2} = q_{k}^{(G)}(j)\ket{j_1, j_2}
\end{equation}
whence the values $q_{k}^{(G)}(j)$ characterize the noisy gate $G$. \revisionB{The amplitude $q_{k}^{(G)}(j)$ describes the transition from basis state $j$ to state $k$ and need not be symmetric.} \revisionA{
We rely on the linearity of the binary noise operator in Eq.~(\ref{eq:binmod}) to transform linear combinations of basis states.
}
\par 
Recall that the probability to measure the outcome $(b_1, b_2')$ of a two-qubit state $\rho$ is given as
\begin{equation}
    \textrm{Prob}(b_1,b_2') =  \Tr \left(\Pi_{1}^{b} \Pi_{2}^{b'} \rho \right)
\end{equation}
where $\Pi_{i}^{b} = \dyad{b_i}{b_i}$ projects onto the value $b$ for register $i$. Hence, measurements in the computational basis estimate the amplitudes $q_{k}^{(G)}(j)$ that characterize the state-dependent binary noise model.
\par
Noise amplitudes for a two-qubit gate may be recovered directly within the binary noise model by preparing and measuring four well-defined inputs. As an example, amplitudes that characterize the identity gate may be measured by preparing the initial state $\rho = \dyad{0_1, 0_2}{0_1, 0_2}$ and observing
\begin{equation}
    |q_{(b_1, b_2)}^{(I)}(0)|^2 = \textrm{Prob}(b_1, b_2')
\end{equation}
Similarly, for $\rho = \dyad{0_1, 1_2}{0_1, 1_2}$,
\begin{equation}
    |q_{(b_1, b_2)}^{(I)}(2)|^2 = \textrm{Prob}(0 \oplus b_1, 1\oplus b_2')
\end{equation}
etc. 
\subsection{Composing State-dependent Noise Models}
Consider two gates, labeled $a$ and $b$, acting in succession on an initial quantum state $\rho_{0}$. Using the above channel model, the prepared states are
\begin{equation}
    \rho_{1} = \xi_{a} (\rho_0) = \sum_{k}{E_{k}^{(a)} U_{a} \rho_0 U_{a}^{\dagger} E_{k}^{(a)\dagger}}
\end{equation}
and
\begin{equation}
    \rho_{2} = \xi_{b} (\rho_1) = \sum_{\ell}{E_{\ell}^{(b)} U_{b} \rho_1 U_{b}^{\dagger} E_{\ell}^{(b)\dagger}}
\end{equation}
or, by substitution,
\begin{equation}
    \rho_2 = \sum_{k, \ell} {E_{\ell}^{(b)}U_{b} E_{k}^{(a)}U_{a} \rho_0 U_{a}^{\dagger} E_{k}^{(a)\dagger} U_{b}^{\dagger} E_{\ell}^{(b)\dagger}}
\end{equation}
If the operator $U_{b}$ is unitary, then
\begin{equation}
    \rho_2 = \sum_{k, \ell} {E_{\ell}^{(b)}U_{b} E_{k}^{(a)}U_{b}^{\dagger} U_{b} U_{a} \rho_0 U_{a}^{\dagger} U_{b}^{\dagger} U_{b}E_{k}^{(a)\dagger} U_{b}^{\dagger} E_{\ell}^{(b)\dagger}}
\end{equation}
and we define
\begin{equation}
    {E'}_{k}^{(a)} = U_{b} E_{k}^{(a)}U_{b}^{\dagger}
\end{equation}
to rewrite
\begin{equation}
\label{eq:comp_rule}
    \rho_2 = \sum_{k, \ell} {E_{\ell}^{(b)} {E'}_{k}^{(a)} U_{b} U_{a} \rho_0 U_{a}^{\dagger} U_{b}^{\dagger} {E'}_{k}^{(a)\dagger} E_{\ell}^{(b)\dagger}}
\end{equation}
This composed channel expresses both noise models as the product of ideal gates $U_{b} U_{a}$ followed by the aggregate noise operator $E_{\ell}^{(b)} {E'}_{k}^{(a)}$.
\subsection{\textsc{CNOT} State-dependent Binary Noise}
We apply the state-dependent binary noise model to the two-qubit \textsc{cnot} gate in which register $q_1$ is the control and $q_2$ is the target. This ideal \textsc{cnot} gate is defined as
\begin{equation}
    U_{\textsc{cnot}}^{(1,2)} = \left(\frac{I_1+Z_1}{2}\right) I_2 + \left(\frac{I_1-Z_1}{2}\right) X_2
\end{equation}
For convenience, we explicitly state the ideal \textsc{cnot} gate with register $q_2$ as the control and $q_1$ as the target as
\begin{equation}
    U_{\textsc{cnot}}^{(2,1)} =  I_1 \left(\frac{I_2+Z_2}{2}\right) + X_1 \left(\frac{I_2-Z_2}{2}\right) 
\end{equation}
Noting that $X_i Z_i = -Z_i X_i$, the above definitions satisfy
\begin{equation}
    U_{\textsc{cnot}}^{(1,2)} X_{1}^{k_1} X_{2}^{k_2} = X_{1}^{k_1} X_{2}^{k_1 \oplus k_2} U_{\textsc{cnot}}^{(1,2)}
\end{equation}
and
\begin{equation}
    U_{\textsc{cnot}}^{(2,1)} X_{1}^{k_1} X_{2}^{k_2} = X_{1}^{k_1\oplus k_2} X_{2}^{k_1} U_{\textsc{cnot}}^{(2,1)}
\end{equation}
From these results, we define the conjugated operators
\begin{align}
    {E'}_{k}^{(\textsc{cnot}(1,2))} & = U_{\textsc{cnot}}^{(2,1)} E_{k}^{(\textsc{cnot}(1,2))}U_{\textsc{cnot}}^{(1,2)\dagger} \\ 
    & = X_{2}^{k_1} U_{\textsc{cnot}}^{(2,1)} \hat{q}_{k}^{(\textsc{cnot}(1,2))} U_{\textsc{cnot}}^{(2,1)\dagger}
\end{align}
which are essential for the composition rule in Eq.~(\ref{eq:comp_rule}). As the amplitude operator is diagonal in the computational basis, the remaining term may be written as
\begin{align}
   \nonumber
   & U_{\textsc{cnot}}^{(2,1)} \hat{q}_{k}^{(\textsc{cnot}(1,2))} U_{\textsc{cnot}}^{(2,1)\dagger}  \\
&    = \sum_{m} {q}_{k}^{(\textsc{cnot}(1,2))}(m) U_{\textsc{cnot}}^{(2,1)} \dyad{m}{m} U_{\textsc{cnot}}^{(2,1)\dagger}
\end{align}
This highlights that the amplitudes are characterized by observing the results of the \textsc{cnot} gate directly in the computational basis.
\subsection{SWAP State-dependent Binary Noise}
Consider the initial state
\begin{equation}
    \rho_0 = \sum_{m,n} {\rho_0(m,n) \dyad{m_1, m_2}{n_1, n_2}}
\end{equation}
with $\rho_0(m,n) = \matrixel{m_1, m_2}{\rho_0}{n_1, n_2}$. We model the state prepared by channel for the noisy \textsc{swap} gate $(s)$ as
\begin{align}
    \rho_1 =  \sum_{k,m,n} & E_{k}^{(s)} U^{(s)} \rho_0(m,n) \\ \nonumber
    & \dyad{m_1, m_2}{n_1, n_2} U^{(s)\dagger} E_{k}^{(s)\dagger}
\end{align}
Reordering the sums and applying the \textsc{swap} gate yields
\begin{equation}
    \rho_1 = \sum_{m,n} \rho_0(m,n) \sum_{k} {E_{k}^{(s)} \dyad{m_2, m_1}{n_2, n_1} E_{k}^{(s)\dagger}}
\end{equation}
For the state-dependent binary channel, we obtain the noisy state as
\begin{align}
    \nonumber 
    \rho_1  = \sum_{k, m,n} & \rho_0(m,n) {X_1^{k_1} X_{2}^{k_2} q_{k}^{(s)}(m_2,m_1)} \\
    & \dyad{m_2, m_1}{n_2, n_1} {X_1^{k_1} X_{2}^{k_2} q_{k}^{(s)}(n_2,n_1)}
\end{align}
from which the transition amplitudes $q^{(s)}_{k}$ may be obtained by measurement in the computational basis. 
\par 
As a simplified example, consider $\rho_{0} = \dyad{1,0}{1,0}$, for which
\begin{align}
    \rho_1 & =  \sum_{k} {p_{k}^{(s)}(2) X_1^{k_1} X_{2}^{k_2}  \dyad{0, 1}{0, 1} X_1^{k_1} X_{2}^{k_2} } \\ \nonumber
    & = \sum_{k} {p_{k}^{(s)}(2)\dyad{k_1,k_2\oplus1}{k_1,k_2\oplus1}}
\end{align}
where $p_{k}^{(s)}(2) = |q_{k}^{(s)}(0,1)|^{2}$ is the probability the state transforms under the $k$-th term of the noise model. The probability is recovered from ideal projective measurement as
\begin{equation}
  \Tr\left(\Pi_{1}^{0} \Pi_{2}^{1} {\rho_2} \right) = p_{0}^{(s)}(2)
\end{equation}
\par
It is notable that the \textsc{swap} operation may be defined as a series of \textsc{cnot} operations acting on the same pair of qubits, i.e.,
\begin{equation}
\label{eq:swapcnot}
    U^{(s)} = U^{(1,2)}_{\textsc{cnot}}U^{(2,1)}_{\textsc{cnot}}U^{(1,2)}_{\textsc{cnot}}
\end{equation}
Consequently, we may approximate the channel for the noisy \textsc{swap}~gate as the composition
\begin{equation}
    \xi_{(s)}(\rho) \approx \xi_{\textsc{cnot}}(\xi_{\textsc{cnot}}(\xi_{\textsc{cnot}}(\rho)))
\end{equation}
From knowledge of the transition amplitudes that characterize the \textsc{cnot} channels, we create a composed channel to represent the action of the \textswap~gate. This approximation becomes exact in the absence of identical, state-dependent dynamics, while the accuracy generally depends on the separability of the channels acting in succession.
\subsection{Noisy Measurement}
Measurement plays an important role in estimating the transition amplitudes, and we extend our model to account for noise in the measurement $(m)$ gate, cf.~Refs.~\cite{Maciejewski2020mitigationofreadout,Geller_2020,PhysRevA.103.042605}. We describe measurement noise using a binary model based on the stochastic sum of projection operators that generate the post-selected state as
\begin{equation}
\label{eq:meas}
    \rho(b,b') = \sum_{\ell} { E_{\ell}^{(m)} \Pi_{1}^{b} \otimes \Pi_{2}^{b'} \rho \Pi_{1}^{b} \otimes \Pi_{2}^{b'}  E_{\ell}^{(m)\dagger} } 
\end{equation}
where again $\Pi_{i}^{b} = \dyad{b_i}{b_i}$. The resulting measurement probability is given as
\begin{equation}
    \textrm{Prob}(b,b') = \Tr\left(\rho(b,b')\right)
\end{equation}
The transition operators $\hat{q}_{\ell}^{(m)}$ for the measurement channel can be characterized in the same manner as the unitary channel, i.e., by measuring the observed transition probabilities with respect to the classical input states. 
\revisionB{
This model can capture correlations between measurement errors, which are often important for accurately recovering the transition amplitudes \cite{Geller_2021,bravyi2021mitigating}.
}
\subsection{Quantum State Tomography}
Quantum state tomography (QST) provides an estimate of the quantum state by reconstructing a numerical representation for the density matrix consistent with experimental observations. While QST is resource intensive and ill-suited for characterizing more than a few qubits, it provides an estimate for the complete representation of the quantum state. Here we use QST to estimate the quantum state prepared by experimental \textswap~programs and use these reconstructions to estimate the prepared state fidelity. 
\par 
As shown in Ref.~\cite{PhysRevLett.108.070502}, the quantum state can be reconstructed by measuring with respect to a complete basis over the Hilbert space. We use tomographic measurements with respect to the Pauli operators to reconstruct a single-qubit state. These measurements generalize to $4^n$ operators generated for measuring an $n$-qubit state. In practice, estimating the expectation value of the Pauli operators requires preparing many identical copies of the state and performing measurement with respect to the operator. We use the reconstructed density matrix $\rho$ to calculate the fidelity with respect to the expected \textsc{swap} output state.
\begin{figure}[ht]
\includegraphics[width=\columnwidth]{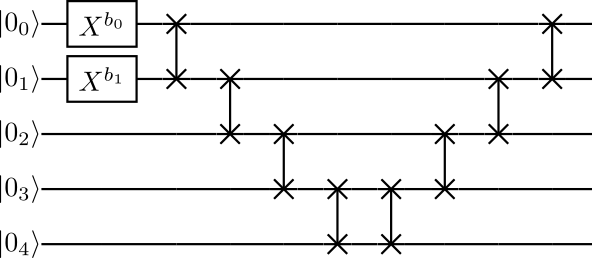}
\caption{A quantum circuit diagram representing a 5-qubit \textsc{swap} program executed on the \textit{boeblingen} device. The cascade of \textsc{swap} operations act on the variable input state defined by the binary pair $(b_0, b_1)$.}
\label{fig:chained_swap_circuit}
\end{figure}
\section{Experimental Results}
\label{sec:test}
We characterized the state-dependent routing dynamics for the experimental quantum processing unit (QPU) \textit{boeblingen} developed by IBM. The \textit{boeblingen} QPU was a 20-qubit register of fixed-frequency superconducting transmons with the hexagonal connectivity shown in Fig.~\ref{fig:boeblingen_graph}. \revisionB{
Each transmon element  encoded computational basis states using the ground and first-excited states labeled as 0 and 1, respectively. In this encoding, the higher energy of the 1 state leads to an asymmetry in the dynamics that include population decay on a time scale $T_1$ \cite{kjaergaard2020superconducting,8681202}.  
}
The \textit{boeblingen} device supported 46 directed connections between register elements with each connection capable of implementing the \textsc{cnot} operation using a sequence of one- and two-qubit gates \cite{chow2011simple}. Similarly, the \textsc{swap} operation was implemented using a sequence of three \textsc{cnot} gates, cf. Eq.~(\ref{eq:swapcnot}). 
\revisionB{Notably, the cross-resonance gate used to generate entanglement between fixed-frequency transmons is known to induce state-dependent $ZZ$ errors due to higher lying levels \cite{chow2011simple,PhysRevLett.125.200504}, offering a physical rationale for state-dependent dynamics. Direct characterization of the state-dependent dynamics, including decoherence, is also possible with current experimental methods \cite{lu2021characterizing}. 
}
\par
We characterized the routing dynamics in \textit{boeblingen} using a set of programs developed with the Qiskit programming library \cite{Qiskit}. Qiskit provided a Python interface to design and execute quantum circuits using an instruction set derived from single-qubit rotation gates and the \textsc{cnot} operation. We estimated the transition amplitudes for the binary noise models described in Sec.~\ref{sec:model} to model the measurement of each register and the available two-qubit gate operations between each available register pair. 
\par
Contemporaneous with these estimates, we evaluated \textsc{swap} programs to move information in the 5-qubit register spanning $q_0$ to $q_4$ as shown in Fig.~\ref{fig:boeblingen_graph}. The remainder of the register elements were not accessed by the test programs. The 5-qubit \textswap~programs were composed manually using the OpenQASM instruction set \cite{cross2017open}, which was then subsequently compiled into the native instructions for the \textit{boeblingen} device. These programs were constrained to express the \textswap~gate as a series of \textsc{cnot} operations while explicit insertions of the barrier command prevented circuit rewriting during compilation. In addition, a custom pass manager method was used to bypass automated circuit rewriting within the Qiskit compiler. 
\par 
Both the characterization and \textswap~programs were performed using multiple initial register states. For the \textswap~programs, we prepared separable pure states of the form
\begin{equation}
\label{eq:init}
    \ket{\Psi(t_0)} = \ket{b^{(0)}_0}\otimes\ket{b^{(1)}_1}\otimes\ket{0_2}\otimes\ket{0_3}\otimes\ket{0_4}
\end{equation}
where the bit values $b^{(0)}$ and $b^{(1)}$ specified the first two register initialization while all other elements are assumed initialized to $0$. Similar initial states at varying register locations were used for characterizing the state-dependent gate noise. For the 5-qubit register, we evaluated the \textsc{swap} program that moves the value in register $q_0$ to register $q_4$ where, in the absence of noise, the expected final state is 
\begin{equation}
    \ket{\Psi(t_5)} = \ket{b^{(1)}_0}\otimes\ket{0_1}\otimes\ket{0_2}\otimes\ket{0_3}\otimes\ket{b^{(0)}_4}
\end{equation}
\par 
Each circuit with fixed input was sampled 8,192 times to gather statistics on the frequency with which specific output states were observed. In total, approximately 7,500 circuits were required for quantum state reconstruction and approximately 50 circuits for heuristic characterization. This amounted to several days of wall time to execute these circuits, including the queue and classical post-processing times. During the course of testing and calibration, the \textit{boeblingen} device underwent automated calibration and tuning procedures to maintain accurate operation of the gates. Our analysis did not account for any intermittent changes in the device itself, a point which we address in our later discussion.
\begin{figure}[t]
    \centering
    \subfloat{
    \includegraphics[width=0.49\textwidth]{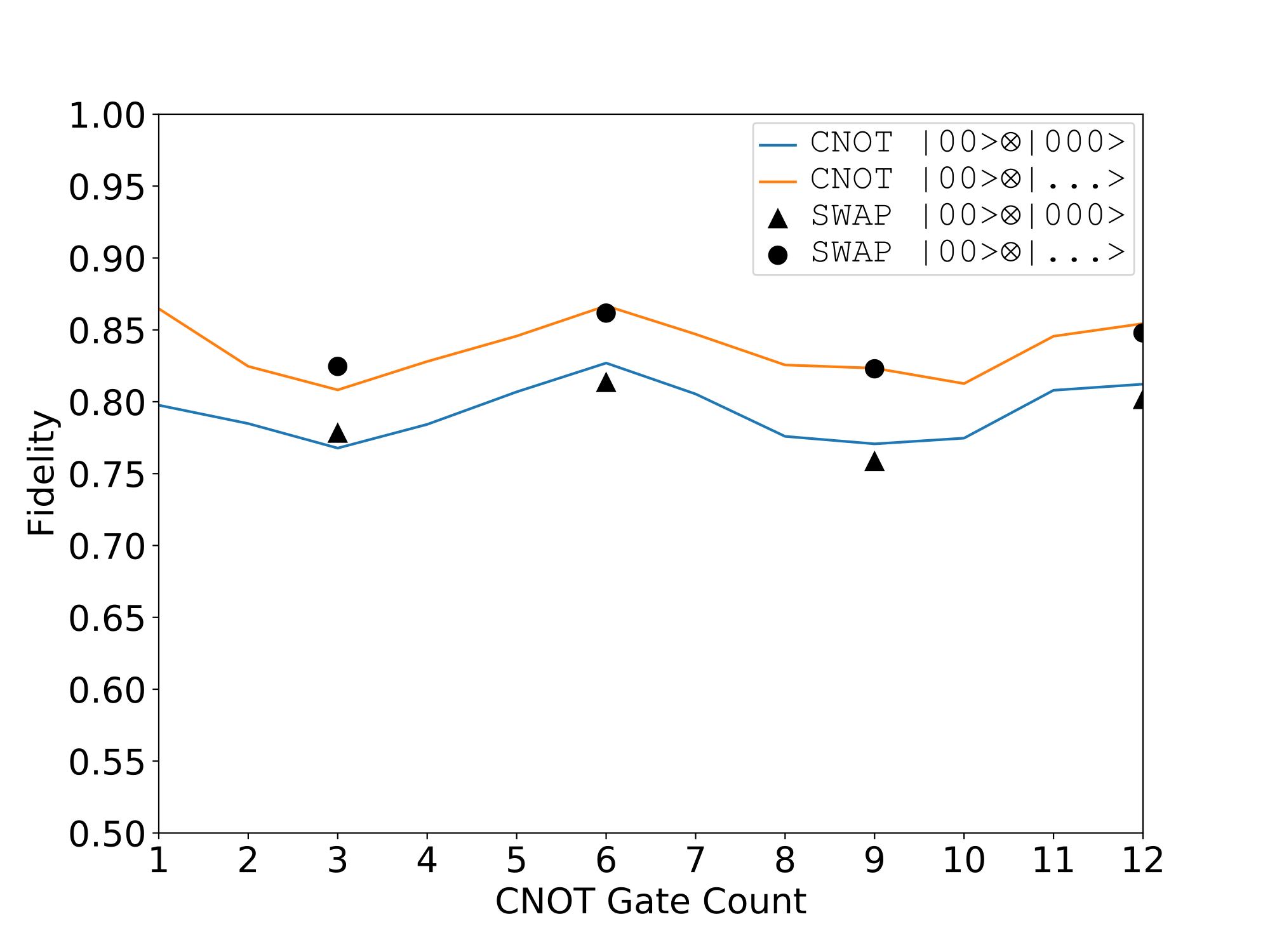}
    }
\\
    \subfloat{
    \includegraphics[width=0.49\textwidth]{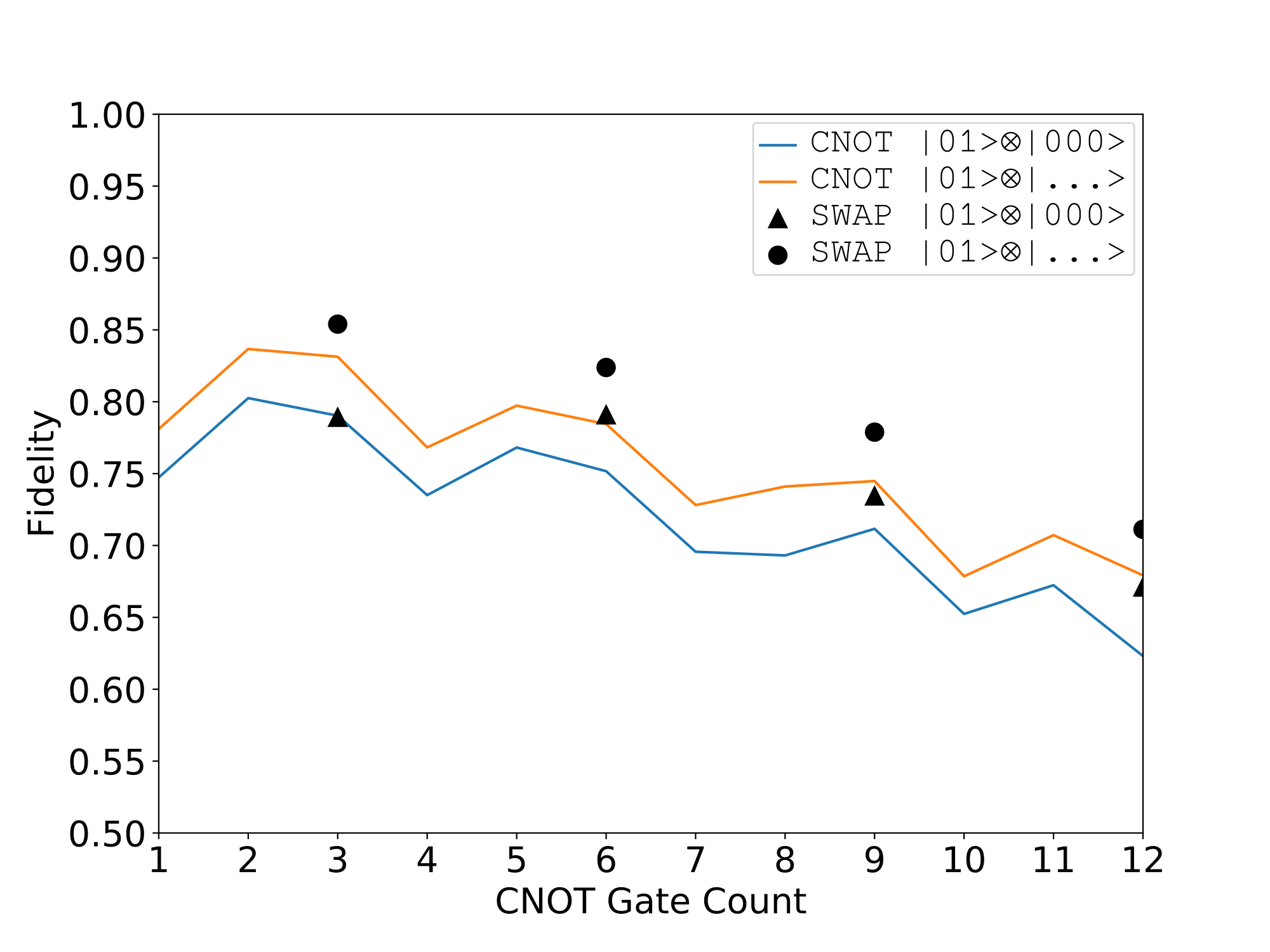}
    }
    \caption{Estimated fidelity of the tomographic reconstructed state with respect to the ideal target state prepared by variable-length \textsc{swap} and \textsc{cnot}~programs. (top) Fidelity for the initial state prepared with $b^{(0)} = b^{(1)} = 0$. (bottom) Fidelity for the initial prepared state with $b^{(0)} = 0$ and $b^{(1)} = 1$. Fidelities correspond to the complete 5-qubit state and the reduced two-qubit state. Points show similar data for the corresponding \textsc{swap} gates.}
    \label{fig:boeblingen_cnot_swap_tomography}
\end{figure}
\subsection{Tomographic Fidelity}
We estimated the fidelity of the quantum state of a 5-qubit register prepared by sequential two-qubit operations using tomographic reconstructions. 
We generated tomographic measurement circuits using the Qiskit Ignis toolkit \cite{Qiskit}
and appended these circuit to a \textsc{swap}~or \textsc{cnot}~program of fixed length. The observed outcomes from these tomographic measurements were then used to perform maximum-likelihood reconstruction with least-squares error \cite{PhysRevLett.108.070502}. We used the built-in Ignis method \texttt{StateTomographyFitter} to perform a convex optimization to reconstruct a density matrix that minimizes the least-square error and, using the reconstructed state, we calculated the fidelity defined by Eq.~(\ref{eq:fidelity}) with respect to the ideal output state expected from the corresponding noiseless program. 
\par
We present estimated fidelities for the quantum states prepared by a series of either \textsc{cnot} or \textsc{swap} operations with different initial states. Figure \ref{fig:boeblingen_cnot_swap_tomography} shows the fidelities for the complete five-qubit state of the register and the reduced two-qubit state of elements $q_0$ and $q_1$ with respect to gate sequence. As expected, we found the estimated fidelity for an $N$-\textsc{swap} program closely matches that of the 3$N$-\textsc{cnot} program for $N = 1$ to $4$. However, the observed behavior and agreement was dependent on the initial state for the program. The case $b_1^{(1)} = 0$ demonstrated an overall higher fidelity and better agreement as compared to the case $b_{1}^{(1)} = 1$. In addition, a steeper decline in fidelity was observed for $b_{1}^{(1)} = 1$, indicating a higher gate error when using this state. 
\par 
In addition, we note the presence of small oscillations in the fidelity with respect to gate sequence as shown in Fig.~\ref{fig:boeblingen_cnot_swap_tomography}. These oscillations vary with the starting state of the register and, again, indicate a state-dependent dynamics present in the error processes. While our compiled programs enforced barriers to prevent unforeseen optimizations of gate operations, the oscillation suggest periodic improvements in either gate fidelity or measurement calibration that remain unexplained. In all cases, the fidelity of the two-qubit state was consistently lower than the five-qubit state. This reduction in fidelity is only weakly dependent on the number of applied gates and suggestive that initial correlations may be present between the registers prior to gate operation. 
\par 
\subsection{Gate Characterization and Parameter Estimation}
\begin{figure}[ht!]
\centering
\includegraphics[width=\linewidth]{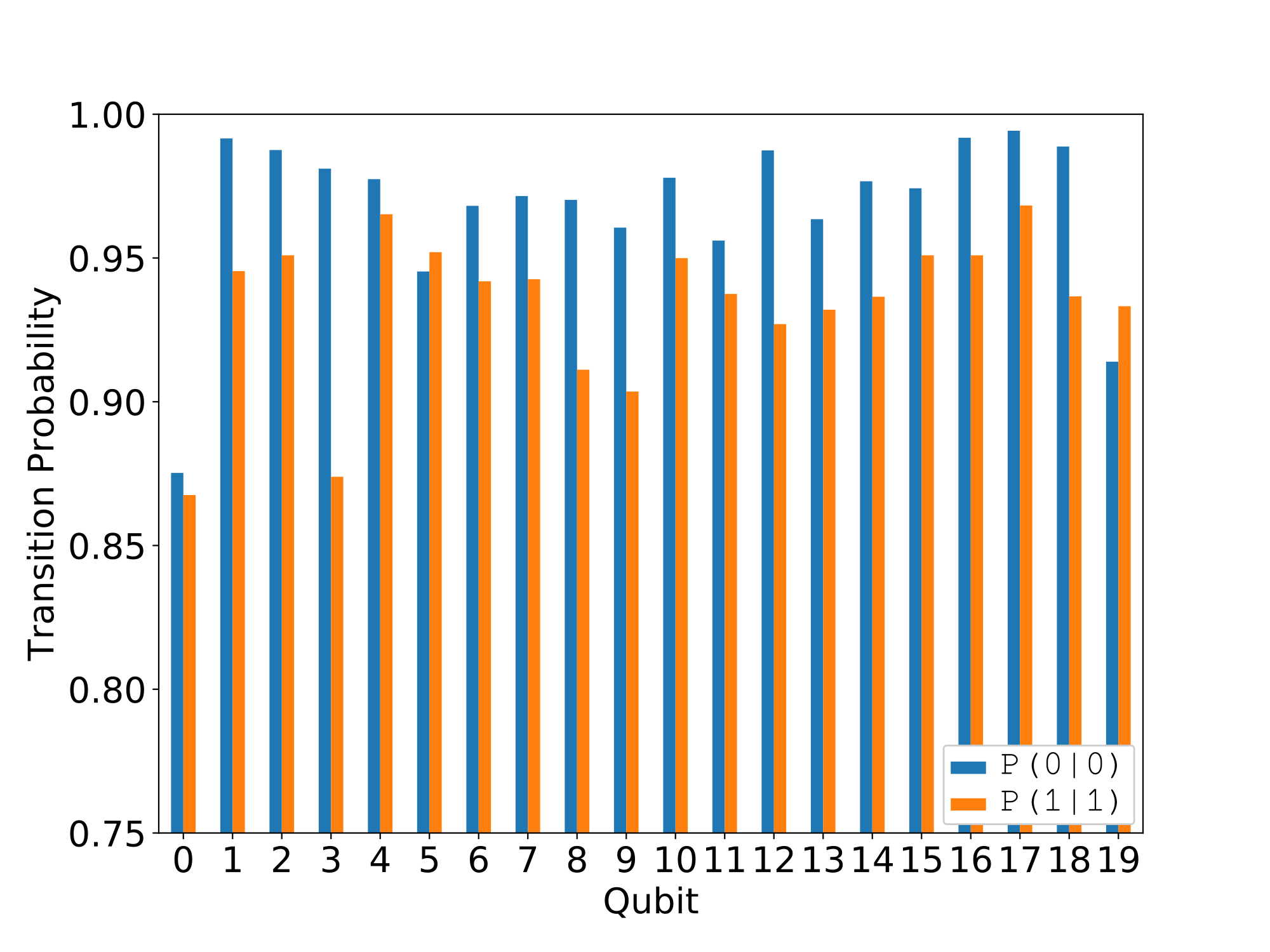}
\caption{State-dependent transition probabilities for measurements on the \textit{boeblingen} device. Each register element is characterized by the probability to observe the expected output state as noted by the conditional probabilities $P(0|0)$ and $P(1|1)$ respectively.}
\label{fig:boeblingen_measurement_characterization}
\end{figure}

Characterization of the state-dependent transition probabilities for measurements on the \textit{boeblingen} device are shown in Fig.~\ref{fig:boeblingen_measurement_characterization}. We observed a range of transition probabilities associated with measurement across different register elements. Moreover, the measurement errors were observed to vary with the expected state and this variation was not consistent across register elements. We fit the state-dependent amplitudes for the measurement model presented in Eq.~(\ref{eq:meas}) using the observed transition probabilities for each qubit. \revisionB{We inverted the resulting linear system of equations to recover the transition matrix using the least-squares method by assuming the single gates used in state preparation were effectively noiseless. This linear fit to state-dependent measurement errors was subsequently used to post-process further characterizations.}
\par
\begin{figure*}[ht]
    \centering
    \subfloat{
      \includegraphics[width=\textwidth]{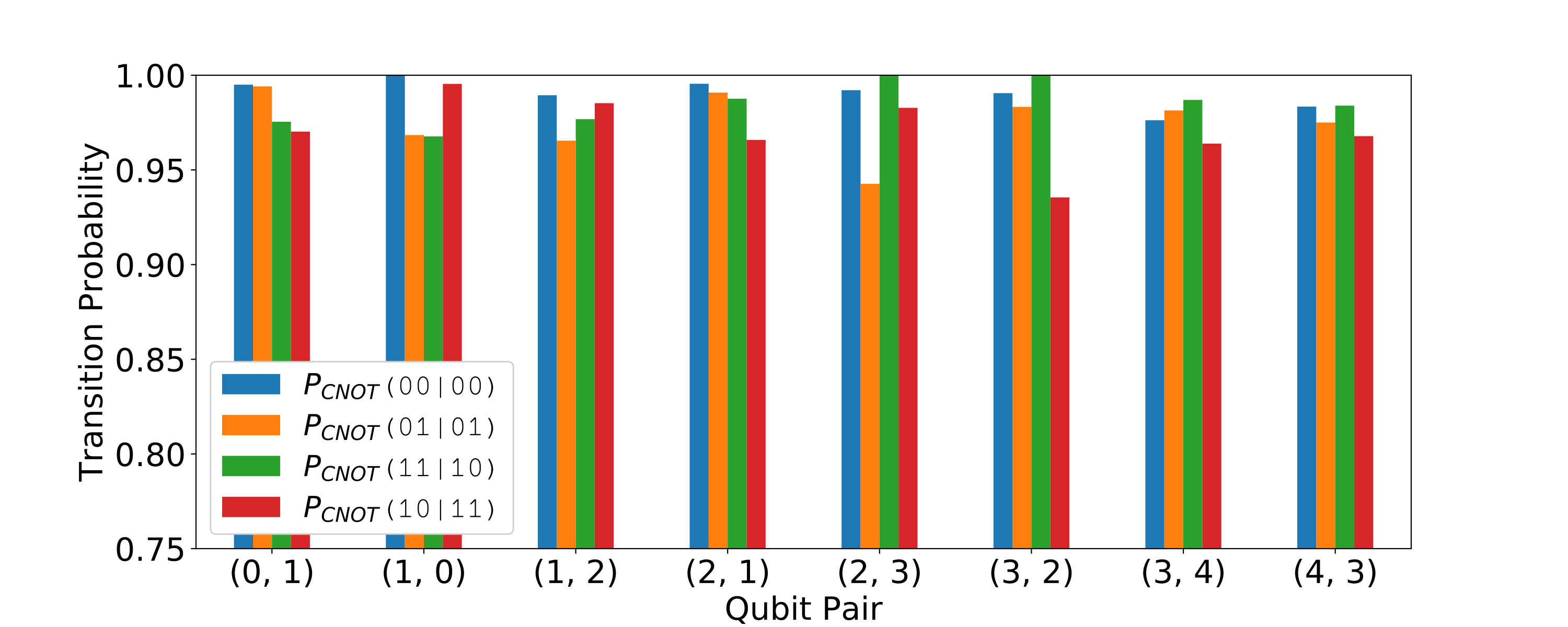}
    }
    \\ 
    \subfloat{
      \includegraphics[width=\textwidth]{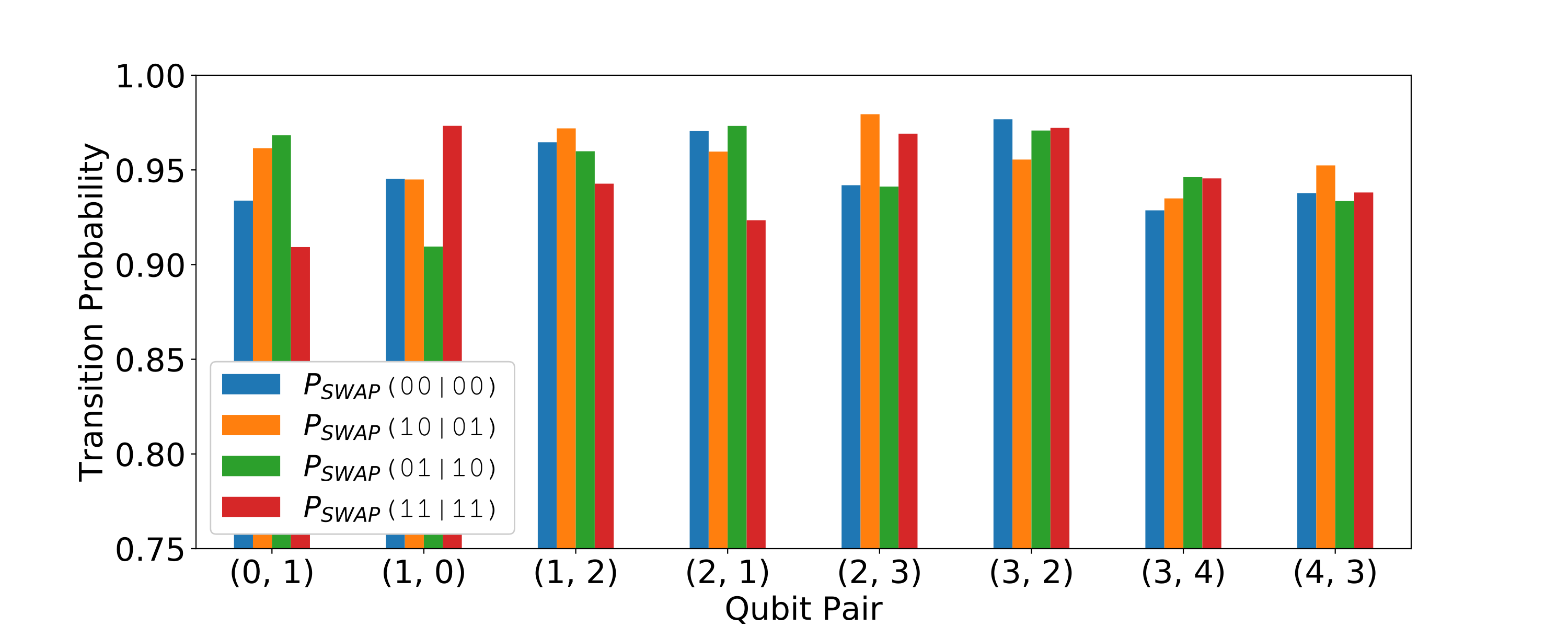}
    }
    \caption{The observed transition probabilities for the (left) \textsc{cnot} and (right) \textsc{swap} operations on the \textit{boeblingen} device. The probability $P(a_i,a_j|b_i,b_j)$ represents the frequency with which the correct state is observed after either the (left) \textsc{cnot} or (right) \textsc{swap} operation given the indicated input state. Deviation from unity indicates a corresponding error rate. We show characterizations for those qubits used in the \textsc{swap} program tested but we collected a complete characterization for the device used.}
    \label{fig:boeblingen_corrected_cnotswap_tp}
\end{figure*}

We next characterized the state-dependent transition probabilities for the \textsc{cnot} and \textsc{swap} gate operations.
Characterization in the computational basis yielded observed transitions that included the influence of noise from the measurement process as well as the gate operation. Noise was removed from the observed probabilities by post-processing with the measurement error model previously characterized. Post-processed observations yielded the state-dependent transition probabilities for the \textsc{cnot} and \textsc{swap} gates. We show the probabilities for each logical transition of these gates in Fig.~\ref{fig:boeblingen_corrected_cnotswap_tp}.
These characterizations provide additional evidence for non-uniform state-dependent transitions amplitudes. While errors vary with respect to register pair as well as the directionality of the gate, we did not identify a generalized  pattern in the amplitudes. \revisionC{In particular, the amplitudes with registers $q_1$ and $q_3$ in Fig.~\ref{fig:boeblingen_corrected_cnotswap_tp} are nominal despite potential cross-talk from the unaddressed registers $q_6$ and $q_8$, respectively. }
\section{Predicted Routing Fidelity}
\label{sec:disc}
We used the experimental characterizations of the noisy gates to fit the state-dependent binary noise model described in Sec.~\ref{sec:model}. We constructed noisy gate models for each \textsc{cnot} and \textsc{swap} gate that were dependent on register pair and directionality, and we used these models to calculate the expected quantum state following a sequence of noisy gates. The fidelity of the simulated noisy quantum state was calculated with respect to the expected noiseless outcome. 
We simulated the fidelity for a \textsc{swap} program with $N = 8$ using noise models for register elements $q_0$ to $q_4$ for four possible input states. 
\par
\begin{figure}
    \centering
    \subfloat{
      \includegraphics[width=0.49\textwidth]{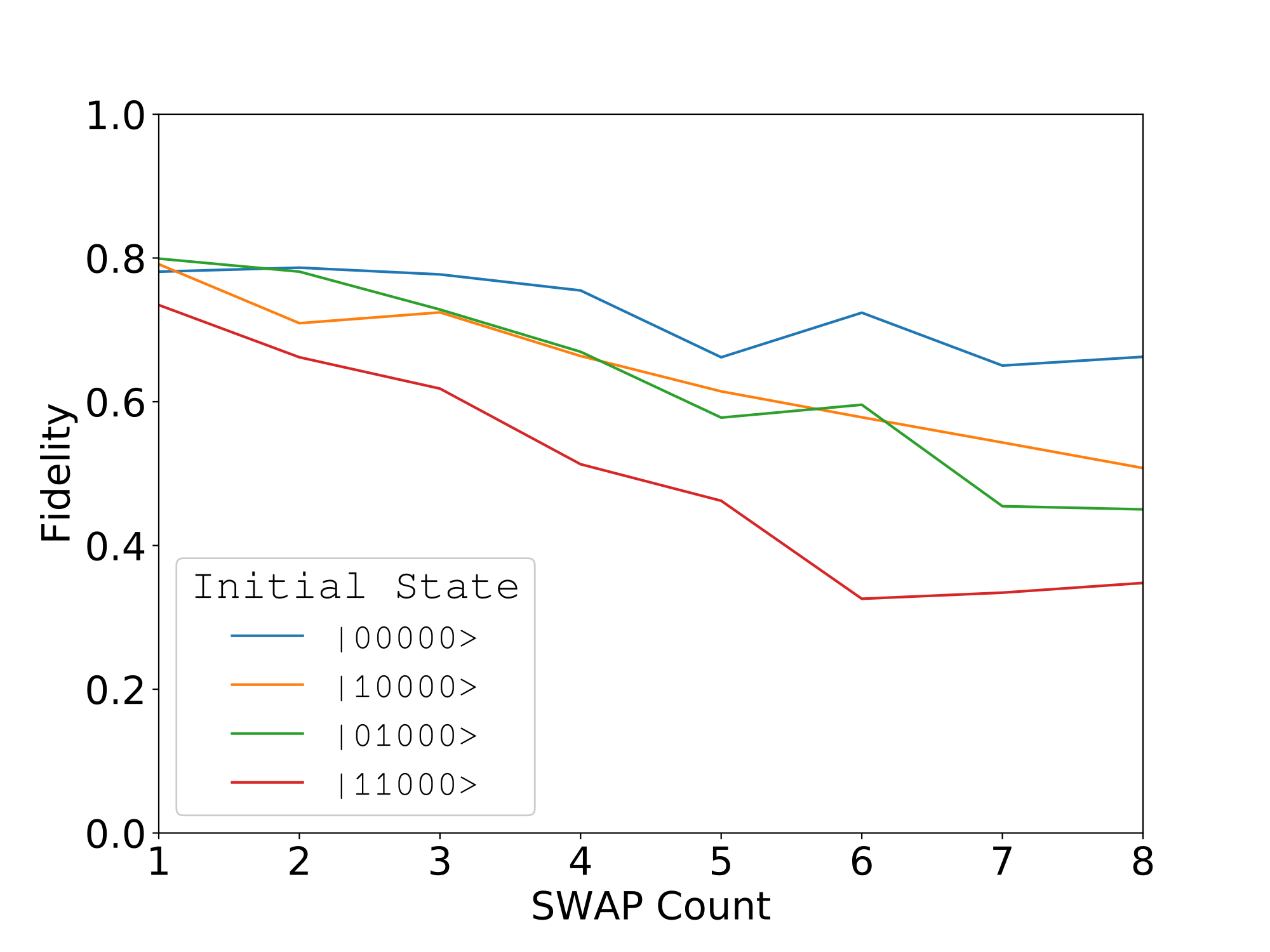}
    }
    \\ 
    \subfloat{
      \includegraphics[width=0.49\textwidth]{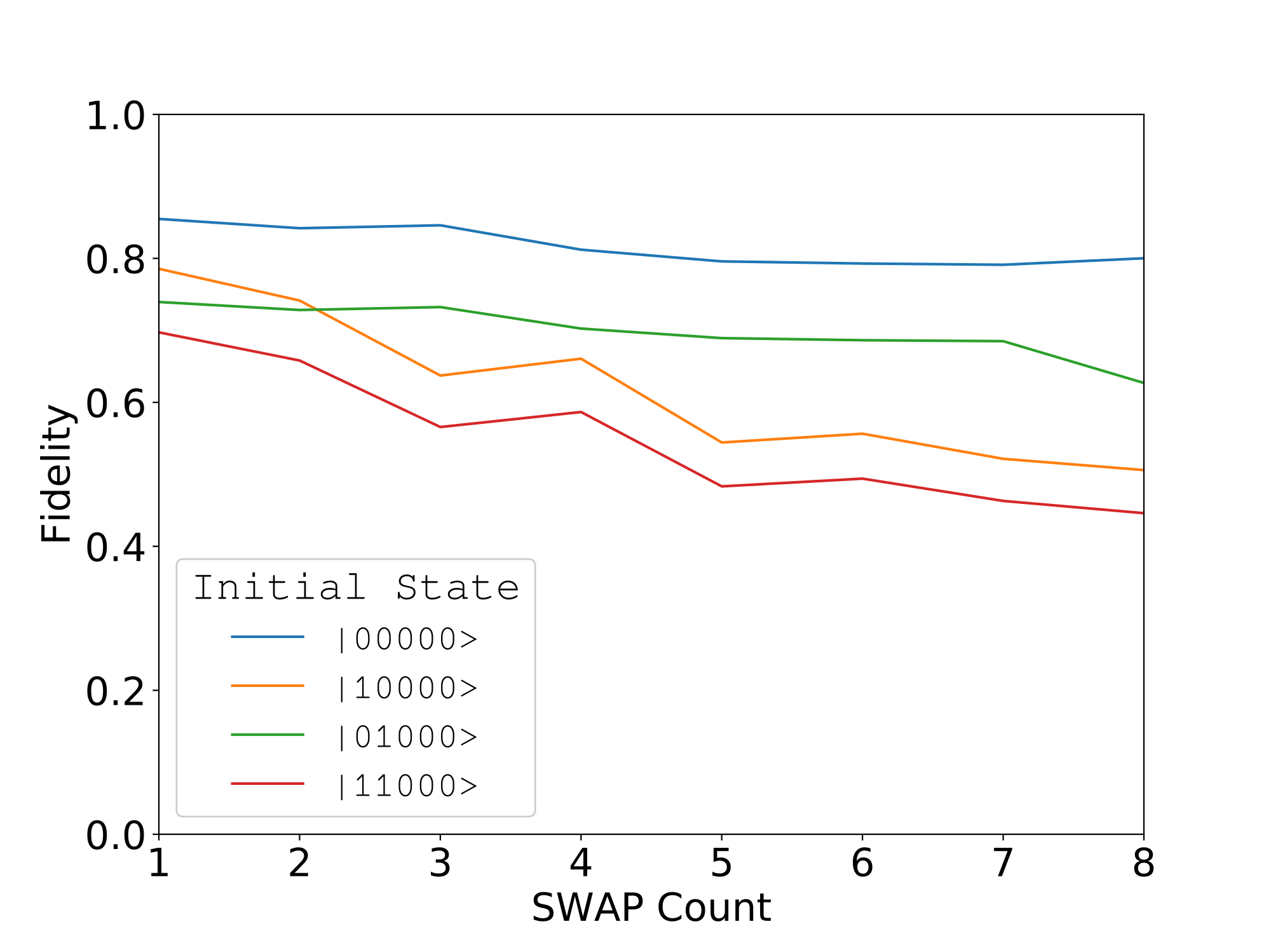}
    }
    \caption{The fidelity from quantum state tomography (top) is compared to the estimated fidelity (bottom) calculated using the composite state-dependent noise model. We plot both fidelities with respect to the number of \textsc{swap} operations for initial states of $\ket{00}$, $\ket{10}$, $\ket{01}$, $\ket{11}$.}
    \label{fig:boeblingen_heuristic_comparison}
\end{figure}

 
\revisionB{
Figure~\ref{fig:boeblingen_heuristic_comparison} presents a comparison of the predicted fidelity using our heuristic model with the observed fidelity recovered from tomographic reconstruction. Each line corresponds to one of the four different input states and the estimated fidelity is plotted with respect to the number of \textsc{swap} gates used for routing across the 5-qubit register on the \textit{boeblingen} device. As shown in the bottom panel of Fig.~\ref{fig:boeblingen_heuristic_comparison}, a clear difference in behavior emerges for the fidelity depending on input state. Starting in the `00' state always yields the highest fidelity ranging from $(1,0.854)$ to $(8,0.800)$, while starting in the `11' state yields the lowest fidelity ranging from $(1,0.697)$ to $(8,0.446)$. Both fidelities generally decrease with \textsc{swap} count with modest fluctuations. The `10' input state is found to mirror the sharper decline in fidelity, ranging from $(1,0.786)$ to $(8,0.506)$, of the `11' state due to the common control bit value. Similarly, the fidelity for the `01' state initial, ranging from $(1,0.739)$ to $(8,0.627)$, approximates that of the `00' fidelity and the corresponding noise dynamics for a  `0' control bit.
}
\par
\revisionB{
By comparison, the top panel of Fig.~\ref{fig:boeblingen_heuristic_comparison} shows the experimentally estimated fidelity. The same trends in behavior are confirmed, namely, the `00' state decays more slowly than the `11' state, the latter has a much lower fidelity, and the `10' and `01' states lie between these extremes. However, the estimated fidelities are not as clearly distinguished. The fidelities range for the `00' state from $(1,0.781)$ to $(8,0.663)$, for the `11' state from $(1,0.735)$ to $(8,0.348)$, for the `01' state from $(1,0.799)$ to $(8,0.450)$, and for the `10' state from $(1,0.791)$ to $(8,0.508)$. With differences of less than $10\%$ at $N=1$ and up to $28\%$ at $N=8$, the model follows the qualitative trends in the fidelity with respect to initial state and length of the \textsc{swap} sequence. 
}
\par
\revisionB{
Some of the quantitative disagreement in the observed and simulated fidelities may be due to fluctuations in the device parameters during  the tomographic measurements \cite{dasgupta2020characterizing}. It is likely that other sources of noise, including accessing higher energy states of the transmons, as well as modeling errors also contribute. For the purposes of routing, however, identifying the relative ordering of the state-dependent fidelities is the necessary information \cite{booth2018comparing,Murali:2019:NCM:3297858.3304075,Tannu:2019:QCE:3297858.3304007,vanmeter2019,10.1117/12.2526670,li2019tackling}. 
} 
State-dependent dynamics impacts a routing algorithm by potentially increasing local edge cost differences resulting in different paths being selected. Only relative differences in edge weights determine the shortest path and if the increase in accuracy is equal for all edges then for actual program execution the algorithm will select the same path. 
We compare this decision process to previous results by Tannu and Querishi that suggested measurement of the $\ket{1}$ state should be avoided when possible due to higher error rates \cite{tannu2019ensemble}. 
\par
Our analysis of the state-dependent routing dynamics in the \textit{boeblingen} device revealed lower fidelities for the  state $\ket{1}$ relative to $\ket{0}$. The observed fidelities provide additional evidence that a state-dependent noise model is more accurate for estimating routing fidelity, as anticipated for transmon devices using the cross-resonance gate \cite{chow2011simple}. 
\revisionA{
An arbitrary superposition of these basis states is expected to exhibit a similarly weighted superposition of these error rates but this prediction was not tested here.
} 
One approach is to average over the asymmetric error rates characterizing paths in a device to yield an average case behavior. This may be implemented by applying alternating rotations to the arbitrary two-qubit state ($X_i X_j$) between successive \textsc{cnot} operations. This is similar to the principle of twirling, which has been used within the context of quantum communication to average the noise of poorly characterized quantum channels \cite{PhysRevLett.76.722}. More generally, twirling has been applied to average gate fidelity \cite{anwar2005practical,PhysRevLett.109.070504,PhysRevA.71.012319}, and it may prove useful as a paradigm for making routing more robust to state-dependent noise. While twirling may average the error within channels, this does not necessarily coincide with an improvement in routing performance and efficiency. 
\par
State tomography could predict the routed path with the best final state fidelity, but it becomes unfeasible due to the required amount of characterization. As a point of comparison, state tomography for the circuit depicted in Fig.~\ref{fig:chained_swap_circuit} required over seven thousand circuits comprised of over 100,000 quantum operations, sampled 8,192 experimental each for a total of over 63 million shots. This took several days of actual time. This was many times longer than the time required by our heuristic characterization, which required less than 50 circuits, comprised of only 75 gates, and only 344,000 shots. \revisionA{This efficiency arises because our binary noise model does not characterize coherent noise, a simplification that may exclude important noise process.} In general, developing additional heuristics will help reduce the resources required while keeping the routed path close to near-optimal \cite{PhysRevA.103.042603}. This will be an important consideration for compiler optimizations in which a target NISQ device has unstable noise characteristics \cite{dasgupta2020characterizing}.
\section{Conclusion}
We have investigated how state-dependent noise affects routing dynamics through the experimental characterization of the routed state fidelity on the IBM \textit{boeblingen} device. Using quantum state tomography, we estimated the register fidelity after variable length \textswap~programs and discovered state-dependent dynamics in the computational basis. We modeled these noisy routing dynamics using  state-dependent binary noise defined in terms of transition amplitudes that characterize the \textsc{cnot} and \textsc{swap} operations. We also showed how a noise model for the \textswap~gate composed from individual characterizations of the \textsc{cnot}~gates compared to the observed fidelity. Notably, the composed model approximates the relative differences in fidelity with respect to the estimated quantum state.
\par
The significance of the state-dependent noise model is two fold. First, by accounting for the state-dependent transition amplitudes, the noise model offer a more accurate representation of the dynamics observed in experiment. Second, the composition of individual, pair-wise models for \textsc{cnot} gates into a model for noisy \textswap~gates greatly simplifies the characterization of the \textswap~program. Both outcomes offer advantages to programming NISQ devices through the development of more accurate and efficient heuristics for fast real-time and compile-time routing decisions.  Our findings indicate that consideration should be given to the compilation requirements for NISQ devices. \revisionA{For example, the results presented here have been limited to tests on a superconducting device but this model may be extended to routing in other technologies, such as ion traps, using the same characterization methods.} Heuristics should be implemented where possible, saving classical pre-processing. These heuristics may be used for real-time routing decisions with uncertainty in predicted noise levels.
\section{Acknowledgments}
The research is supported by the Department of Energy Office of Science Early Career Research Program. This work used resources of the Oak Ridge Leadership Computing Facility, which is a DOE Office of Science User Facility supported under Contract DE-AC05-00OR22725. 

\bibliographystyle{ieeetr}
\bibliography{refs}
\end{document}